\documentclass{PoS}

\PoS{PoS(LAT2005)051}

\title{Maximal twist and the spectrum of quenched twisted mass lattice QCD }

\ShortTitle{Maximal twist and the spectrum of quenched twisted mass lattice QCD
}

\author{\speaker{Abdou M. Abdel-Rehim}\\
        Department of Physics, University of Regina, Regina, SK,
        Canada, S4S 0A2\\
        E-mail: \email{rehimhaa@uregina.ca}}

\author{Randy Lewis\\
        Department of Physics, University of Regina, Regina, SK,
        Canada, S4S 0A2}
     
\author{R. M. Woloshyn\\
        TRIUMF, 4004 Wesbrook Mall, Vancouver, BC,
        Canada, V6T 2A3}

\abstract{Results on the hadron masses for a degenerate doublet
          of up and down quarks from quenched twisted mass lattice QCD
          at maximal twist are presented. Two definitions of maximal
          twist are used and the hadron masses for these definitions
          are compared. Mass splittings within the $\Delta(1232)$
          multiplet due to flavor breaking effects are discussed.}

\FullConference{XXIIIrd International Symposium on Lattice Field Theory\\
		 25-30 July 2005\\
		 Trinity College, Dublin, Ireland}

\begin{document}

\section{Introduction}
With twisted mass lattice QCD(tmQCD) \cite{Frezzotti:2000nk}, it
became possible to perform lattice simulations with Wilson type quarks
at small quark masses. This is because of the fact that
the Wilson Dirac operator with a twisted mass term is free of
unphysical zero eigenmodes as long as the quark mass and the twist
angle are nonzero. Another extremely useful property of tmQCD is that
when the twist angle equals $\frac{\pi}{2}$ (maximal twist),
discretization errors of $O(a^{2k+1});k=0,1,2,\dots$ of correlation
functions of gauge invariant, multiplicatively renormalizable, multi-local operators
are either automatically absent or could be removed by taking a simple
averaging procedure\cite{Frezzotti:2003ni}. This is particularly simple
in the case of extracting masses from 2-point correlators, since the
improvement will be automatic and no averaging is needed. These
features make tmQCD at maximal twist an attractive method to study
the hadronic spectrum. An important element in this program is the
definition of maximal twist and how the uncertainty in this definition
could affect the expected reduction of discretization errors.  
Initial quenched simulations \cite{Jansen:2003ir} used the condition of
maximal twist from the vanishing of the pion mass as computed in the untwisted action
(Wilson maximal twist). Improved results were obtained \cite{Jansen:2003ir}\cite{Abdel-Rehim:2004gx}
using Wilson maximal twist at not so light quark masses. 
However, for very light quark masses
(close to the interesting physical up and down quark masses), undesired discretization effects became
evident\cite{Bietenholz:2004wv}\cite{Abdel-Rehim:2005gz}. In addition tmQCD chiral perturbation
theory \cite{Aoki:2004ta}\cite{Sharpe:2004ps} as well as a study of discretization errors in tmQCD
\`a la Symanzik \cite{Frezzotti:2005gi} argued for the need for an optimal determination 
of the maximal twist condition in the light quark mass regime. Improved definitions
of maximal twist \cite{Farchioni:2004fs}\cite{Sharpe:2004ny}
have been used in recent simulations \cite{Abdel-Rehim:2005gz}\cite{Jansen:2005gf}\cite{Farchioni:2004fs},
using PCAC or the vanishing of wrong parity matrix elements(parity maximal twist).
For more discussion, see \cite{Sharpe:2005rq}.

In this work, results for the quenched hadron spectrum from a degenerate doublet of up and down quarks are presented
for various quark masses and at different lattice spacings. Simulations were performed implementing both
the Wilson maximal twist and parity maximal twist definitions and results are compared.

\section{Action at Maximal Twist}
The action is given by
\begin{equation}
S[\psi,\bar{\psi},U,\beta]=S_F[\psi,\bar{\psi},U]+S_g[\beta,U],
\end{equation}
where $S_g$ is the standard Wilson plaquette action for the gauge field at coupling $\beta$, while
$S_F$ is the tmQCD action for a degenerate doublet $\psi$ of the up and down quarks. In the twisted
basis
\begin{equation}
S_F[\psi,\bar{\psi},U]=a^4\sum_x\bar\psi(x)\left(M+i\mu_q\gamma_5\tau^3
     +{D\!\!\!\!\!/}_{\rm{W}}\right)\psi(x),
\end{equation}
where
\[{D\!\!\!\!\!/}_{\rm{W}}=\frac{1}{2}\sum_\nu\left(\gamma_\nu\nabla_\nu+\gamma_\nu\nabla_\nu^*
     -a\nabla_\nu^*\nabla_\nu\right),\]
\[\nabla_\nu\psi(x) = U_\nu(x)\psi(x+a\hat\nu) - \psi(x),\quad\quad \nabla_\nu^*\psi(x) = \psi(x) - U_\nu^\dagger(x-a\hat\nu)\psi(x-a\hat\nu).\]
When the mass parameter $M=M_{critical}$, we are at maximal twist and the bare quark mass equals $a\mu_q$. Simulations 
are performed with $M_{critical}$ defined by the value that corresponds to a massless pion in the standard Wilson
action (Wilson maximal twist)\cite{Jansen:2003ir}, and with $M_{critical}$ defined by the value when the wrong parity mixing
between the physical vector and pseudoscalar vanishes (parity maximal twist), i.e.
\begin{equation}
\sum_{\vec{x}}\left<V_\nu^-(\vec{x},t)P^+(0,0)\right>=0.
\label{parity-definition}
\end{equation}
\section{Simulations}
\begin{table}
\begin{tabular}{cccccl}
~~~$\beta$~~~ & ~~~\#sites~~~ & \#configurations &~~~$aM_{critical}$~~~&~~~$a\mu_q$~~~&~~~$am_q^{physical}$  \\
\hline             
5.85 & 20$^3\times40$ & 1000  & -0.9071 & 0.0376     &$\sim m_s$   \\
     &                &       &         & 0.0188     &$\sim m_s/2$ \\
     &                &       &         & 0.012527   &$\sim m_s/3$ \\
     &                &       &         & 0.00627    &$\sim m_s/6$ \\
\hline
     &                & 600   & -0.8965 & 0.0376     &$\sim m_s$\\
     &                &       & -0.0971 & 0.0188     &$\sim m_s/2$\\
     &                &       & -0.9110 & 0.01252    &$\sim m_s/3$\\
     &                &       & -0.9150 & 0.00627    &$\sim m_s/6$\\
\hline
6.0  & 20$^3\times48$ & 1000  & -0.8135 & 0.030      &$\sim m_s$\\
     &                &       &         & 0.015      &$\sim m_s/2$\\
     &                &       &         & 0.010      &$\sim m_s/3$\\
     &                &       &         &0.005       &$\sim m_s/6$\\
\hline
     &                & 600   & -0.8110 &0.030       &$\sim m_s$\\
     &                &       & -0.8170 &0.015       &$\sim m_s/2$\\
     &                &       & -0.8195 &0.010       &$\sim m_s/3$\\
     &                &       & -0.8210 &0.005       &$\sim m_s/6$\\
\hline
6.2  & 28$^3\times56$ & 200   & -0.7363 & 0.02165&$\sim m_s$\\
     &                &       &         & 0.01083&$\sim m_s/2$ \\
     &                &       &         & 0.00722&$\sim m_s/3$\\
     &                &       &         & 0.00361&$\sim m_s/6$\\
\hline
\end{tabular}
\caption{Parameters of the simulations. The bare quark mass $a\mu_q$ values 
were chosen such that the physical quark mass $am_q^{physical}$ corresponds to 
specific fractions of the strange quark mass. For Wilson maximal twist,
the value of $aM_{critical}$ is independent of the bare quark mass value.}
\label{sim_par}
\end{table} 
In Table \ref{sim_par}, the simulation parameters are listed. The values of $M_{critical}$ for the parity definition
of maximal twist were tuned at each $\mu_q$ such that the condition in Eq.\ref{parity-definition} is satisfied.
Standard local interpolating fields for the charged pion, the charged $\rho$ ,
$J^P=(\frac{1}{2})^\pm$ baryons and $J^P=(\frac{3}{2})^\pm$ baryons are used to obtain
the ground state masses\cite{Abdel-Rehim:2005gz}. Using the PCAC relation, the charged pion
decay constant can be obtained from
\begin{equation}
f_\pi=\frac{2\mu_q}{m_\pi^2}|\left<0|P^\pm|\pi\right>|.
\end{equation}
\section{Results}
\begin{figure}
\begin{center}
\includegraphics[width=12cm,height=6cm]{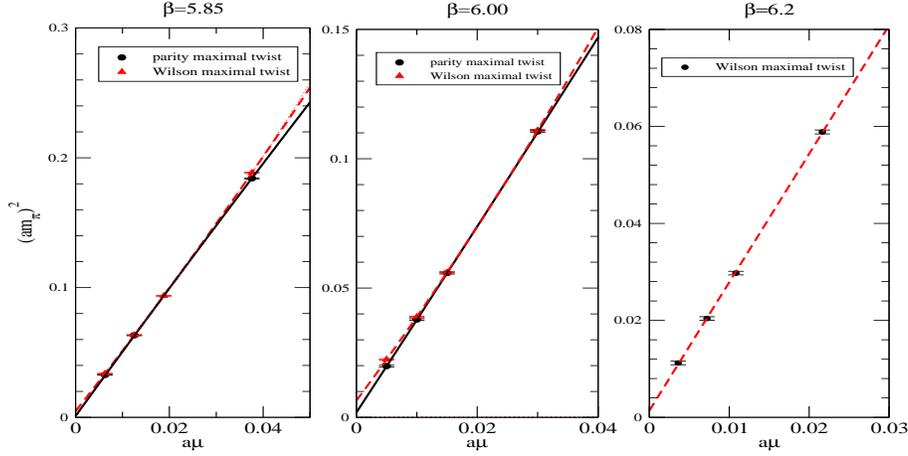}
\caption{The charged pion mass squared (in lattice units) as a function of the 
bare quark mass $a\mu_q$.}
\label{mpivsmu}
\end{center}
\end{figure}

\begin{figure}
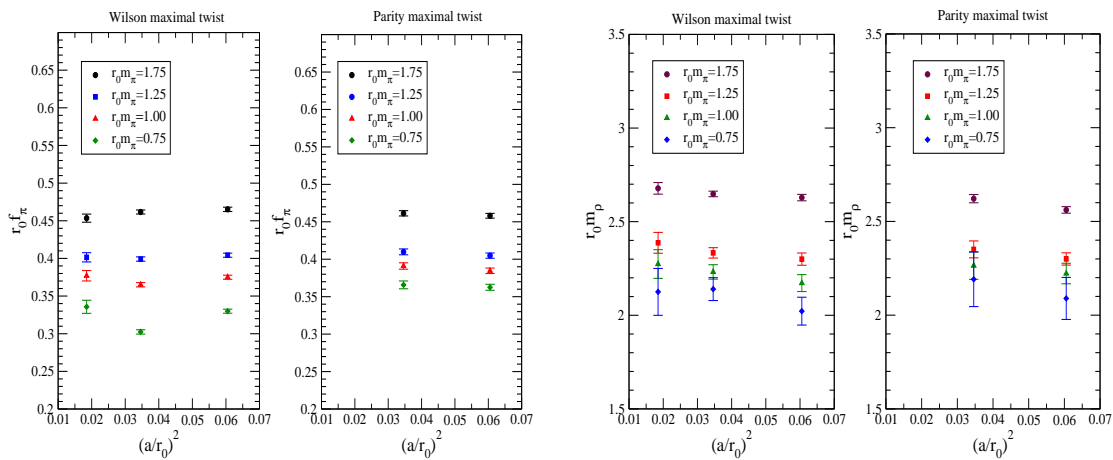

\begin{center}
\includegraphics[width=7cm,height=6cm]{fpiscaling2.eps}
\hspace{0.5cm}
\includegraphics[width=7cm,height=6cm]{mrhoscaling.eps}
\caption{Scaling of the charged pion decay constant and the charged rho mass.}
\label{scaling}
\end{center}
\end{figure}

\begin{figure}
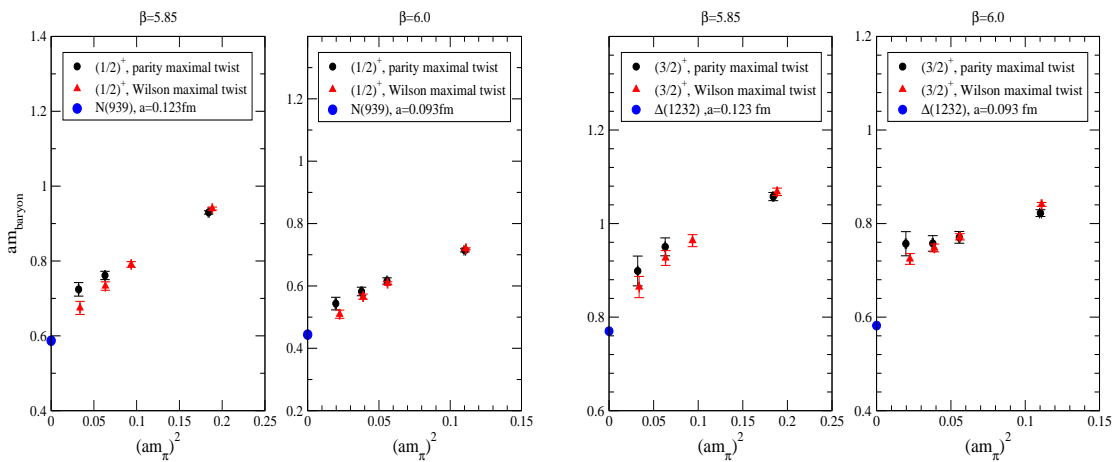

\begin{center}
\includegraphics[width=7cm,height=6cm]{mNvsmpiparity-plus.eps}
\hspace{0.5cm}
\includegraphics[width=7cm,height=6cm]{mDvsmpiparity-plus.eps}
\caption{Baryon mass for $(\frac{1}{2})^+$ and $(\frac{3}{2})^+$ states as a function
of the pion mass squared. The physical nucleon N(939) and $\Delta(1232)$ masses correspond
to the lattice spacings in the legends.}
\label{baryon_mass}
\end{center}
\end{figure}

In Figure \ref{mpivsmu}, the charged pion mass squared is plotted as a function of the bare quark mass. The parity 
definition of maximal twist shows the expected linear dependence at light quark masses with $am_\pi$ approximately
vanishing in the chiral limit. A better scaling behavior of $f_\pi$ is also observed with the parity definition 
of maximal twist as
shown in Figure \ref{scaling}, while the scaling of the rho mass is equally good in both definitions if one takes into 
account the large statistical errors. For the rho mass, results extracted using the $4^{th}$ component of the tensor
operator are also shown which were found, in some cases, to have small statistical errors. In Figure \ref{baryon_mass}, results for
the Nucleon and Delta masses are shown for both definitions of maximal twist. These are about $10\%$ higher than the 
physical values, a result that might be expected due to quenching. 

\begin{figure}
\begin{center}
\includegraphics[width=12cm,height=6cm]{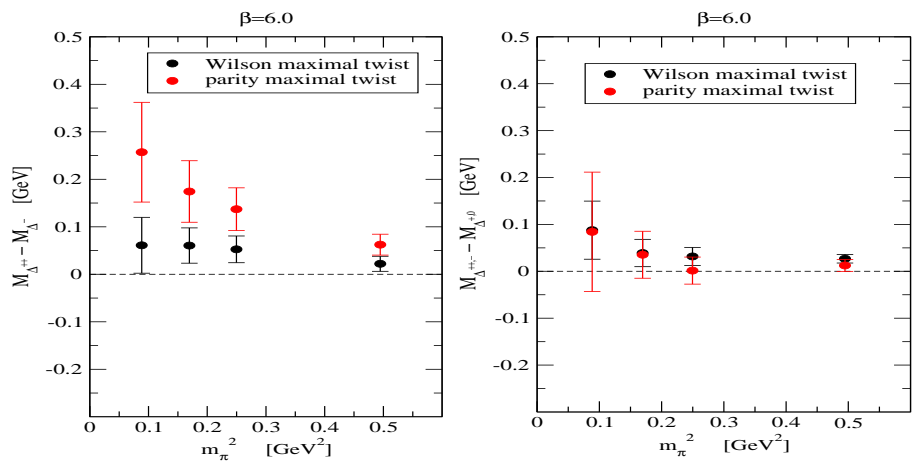}
\caption{$\Delta$ mass splittings.}
\label{Delta_splitting}
\end{center}
\end{figure}

The twisted mass action includes  flavor breaking effects due to the Wilson term. These could be seen as a mass splitting
among isospin multiplets. In case of the $\Delta$ multiplet, there are no disconnected diagrams and one can study the
mass splitting among the four $\Delta$ states. On the average, for a degenerate doublet of the up and down quarks, the
up quark propagator $U(x,y)$ and the down quark propagator $D(x,y)$ are not equal but  are related by \cite{Abdel-Rehim:2005gz}
\begin{equation}
U(x,y)=\gamma_5D(y,x)^\dagger\gamma_5.
\end{equation}
Using this relation, one expects the following mass relation among the $\Delta$ states,
\begin{equation}
(M_{\Delta^{++}}=M_{\Delta^-}=M_{\Delta^{++,-}}) \neq (M_{\Delta^+}=M_{\Delta^0}=M_{\Delta^{+,0}}).
\end{equation}
In Figure \ref{Delta_splitting}, the measured mass splittings are shown for $\beta=6.0$. From these results, one concludes that
there could be a hint of flavor breaking, however, these are hidden by large statistical errors. The flavor breaking effects 
between the charged and neutral pion need the computation of disconnected diagrams which has been calculated in  \cite{Farchioni:2005hf}
\cite{Jansen:2005cg}. A recent study of tmQCD with a strange quark also includes some discussion of flavor breaking\cite{Randy:spectrum_strange}.

\section{Conclusions}
Improved results for the hadron spectrum could be obtained from tmQCD and light quark masses. The parity definition of maximal
twist was found to have a better scaling and chiral behaviour than the Wilson definition of maximal twist. 
\section{Acknowledgements}
This work was supported in part by the Natural Sciences and Engineering Research Council of Canada, the Canada Foundation
for Innovation, the Canada Research Chairs Program and the Government of Saskatchewan.

\providecommand{\href}[2]{#2}\begingroup\raggedright


\begin{thebibliography}{99}
  \bibitem{Frezzotti:2000nk} 
  [Alpha collaboration]R.~Frezzotti, P.~A.~Grassi, S.~Sint and P.~Weisz,
  JHEP {\bf 0108} (2001) 058
  [\href{http://arxiv.org/abs/hep-lat/0101001}{\tt hep-lat/0101001}].

  \bibitem{Frezzotti:2003ni}
  R.~Frezzotti and G.~C.~Rossi,
  JHEP {\bf 0408}, 007 (2004)
  [\href{http://arxiv.org/abs/hep-lat/0306014}{\tt hep-lat/0306014}].

\bibitem{Jansen:2003ir}[XLF Collaboration]
  K.~Jansen, A.~Shindler, C.~Urbach and I.~Wetzorke,
  Phys.\ Lett.\ B {\bf 586}, 432 (2004)
  [\href{http://arxiv.org/abs/hep-lat/0312013}{\tt hep-lat/0312013}].

\bibitem{Abdel-Rehim:2004gx}
  A.~M.~Abdel-Rehim and R.~Lewis,
  Phys.\ Rev.\ D {\bf 71}, 014503 (2005)
  [\href{http://arxiv.org/abs/hep-lat/0410047}{\tt hep-lat/0410047}].

\bibitem{Bietenholz:2004wv}[XLF Collaboration]
  W.~Bietenholz {\it et al.},
  JHEP {\bf 0412}, 044 (2004)
  [\href{http://arxiv.org/abs/hep-lat/0411001}{\tt hep-lat/0411001}].

\bibitem{Abdel-Rehim:2005gz}
  A.~M.~Abdel-Rehim, R.~Lewis and R.~M.~Woloshyn,
  Phys.\ Rev.\ D {\bf 71}, 094505 (2005)
  [\href{http://arxiv.org/abs/hep-lat/0503007}{\tt hep-lat/0503007}].

\bibitem{Aoki:2004ta}
  S.~Aoki and O.~B\"ar,
  Phys.\ Rev.\ D {\bf 70}, 116011 (2004)
  [\href{http://arxiv.org/abs/hep-lat/0409006}{\tt hep-lat/0409006}].

\bibitem{Sharpe:2004ps}
  S.~R.~Sharpe and J.~M.~S.~Wu,
  Phys.\ Rev.\ D {\bf 70}, 094029 (2004)
  [\href{http://arxiv.org/abs/hep-lat/0407025}{\tt hep-lat/0407025}].

\bibitem{Frezzotti:2005gi}
  R.~Frezzotti, G.~Martinelli, M.~Papinutto and G.~C.~Rossi,
  [\href{http://arxiv.org/abs/hep-lat/0503034}{\tt hep-lat/0503034}].

\bibitem{Jansen:2005gf}
  [XLF Collaboration],
  K.~Jansen, M.~Papinutto, A.~Shindler, C.~Urbach and I.~Wetzorke,
  Phys.\ Lett.\ B {\bf 619}, 184 (2005)
  [\href{http://arxiv.org/abs/hep-lat/0503031}{\tt hep-lat/0503031}].

\bibitem{Farchioni:2004fs}
  F.~Farchioni {\it et al.},
  Eur.\ Phys.\ J.\ C {\bf 42}, 73 (2005)
  [\href{http://arxiv.org/abs/hep-lat/0410031}{\tt hep-lat/0410031}].

\bibitem{Sharpe:2004ny}
  S.~R.~Sharpe and J.~M.~S.~Wu,
  Phys.\ Rev.\ D {\bf 71}, 074501 (2005)
  [\href{http://arxiv.org/abs/hep-lat/0411021}{\tt hep-lat/0411021}].

\bibitem{Sharpe:2005rq}
  S.~R.~Sharpe,
  [\href{http://arxiv.org/abs/hep-lat/0509009}{\tt hep-lat/0509009}].

\bibitem{Farchioni:2005hf}
  F.~Farchioni {\it et al.},
  [\href{http://arxiv.org/abs/hep-lat/0509036}{\tt hep-lat/0509036}].

\bibitem{Jansen:2005cg}[XLF Collaboration]
  K.~Jansen {\it et al.}  ,
  [\href{http://arxiv.org/abs/hep-lat/0507032}{\tt hep-lat/0507032}].

\bibitem{Randy:spectrum_strange} 
Abdou M. Abdel-Rehim, Randy Lewis, and R. M. Woloshyn,
 [\href{http://arxiv.org/abs/hep-lat/0509056}{\tt hep-lat/0509056 }].

\end{thebibliography}
\end{document}